# Effects of weak measurements on sudden change of quantum discord

Mei Bai, Xiang-Yu Qin, and Xue-Qun Yan[†]

School of Physical Science and Technology, TianGong University, Tianjin 300387, China

We study the peculiar dynamics of two qubits quantum discord with a sudden change in behavior. It is widely agreed in the literature nowadays that environment induce sudden change behavior of quantum discord. We explore the sudden change phenomenon under the action of local-independent environments, considering the discord of two-qubit which is obtained by means of the weak measurements on one of the two-qubit. It is shown that the sudden change phenomenon does not appear under weak measurements. We find that the quantum discord with weak measurements decays in a monotonic fashion, and show that quantum discord might exhibit a sudden change only for the strong (projective) measurement. We believe that environment is not the main role of the sudden change, but the quantum correlation itself is.

## I.  INTRODUCTION

Quantum discord is recognized as an important resource for offering computational speed-up in quantum information processing tasks [1,2]. It is well studied in various works in recent years, and reveals many new properties in quantum correlations [3-12]. For instance, investigations in the recent past have revealed that quantum discord is a fundamental resource for deterministic one-qubit quantum computation and remote state preparation [13]. Relations of quantum discord with quantum communication have also been pointed out, such as in local broadcasting [14] and quantum state merging [15]. Some authors studied quantum discord in the context of channel discrimination and state discrimination, and conjectured that quantum discord is the necessary ingredient [16]. In addition, quantum discord has also been used in studies of quantum phase transition [17], relativistic effect [18], and biological systems [19]. Among these, however, the exact role played by the quantum discord is unclear.

Realistic quantum systems are open, and their correlations with the surrounding environments are of statistical nature and unavoidable. The inevitable interaction between a quantum system and its environment would destroy coherence and quantum correlations in the system and lead to the reduction of a useful resource. Strategies for preventing the decay of quantum correlations and thus to fight decoherence are important for effective realization of quantum information processing tasks. To explore this issue, understanding the dynamics of quantum correlations in open quantum systems is essential for quantum information processing and quantum computation. Therefore, it is interesting subject to investigate the dynamics of different kinds of correlations in noisy environments. The knowledge of the dynamic behavior of correlations will also help us design suitable protocols to protect correlations during processing. Many efforts have been devoted to the investigate of the dynamics of quantum correlation under various decoherence channels. Those previous results suggest that the dynamical behavior of correlations presented in a composite open quantum system strongly depends on the noise produced by the surrounding environments [20].

---

[†]E-mail: yanxuequn@tiangong.edu.cn；xqyan867@tom.com

An interesting feature of the dynamics of quantum discord is the existence of a sudden change. Some studies have been carried out in dynamical context and found that the dynamics of quantum discord can occur a sudden change phenomenon under different noisy channels [21-24] and maintain a constant value in the long-time limit even when entanglement suddenly disappears. In [25], the peculiar aspect of the dynamics of quantum discord was studied in the presence of non-dissipative decoherence, and the phenomenon of the sudden transition in a finite time interval was reported. Moreover, some of these phenomena have been observed in the recent experiment. Under the action of a real thermal noise environment at room temperature, the dynamics of correlations were investigated in an NMR quadrupolar system [26]; their results confirm that the phenomena predicted in Ref. [21] for phase environments take place even in the presence of an additional thermal noise. The experimental observation exhibited an excellent agreement with the given results of Ref. [27].

On the other hand, as is well known that the seminal paper in which Ollivier and Zurek introduced quantum discord is by discussing the disturbance caused on one subsystem of a bipartite state due to projective measurements performed on the other [28]. According to traditional interpretation of quantum mechanics, measurement of the quantum system collapses it into a new state that is different from the one presented before the measurement. Even for non-disturbing measurements, in which the system variable is unaffected by the measurement, it is not mean that the system state is unaffected by the measurement. Thus, the measurement in quantum mechanics is usually a procedure that destroys many properties of the system. However, weak measurements were introduced by Aharonov and collaborators as a theoretical scheme [29], allowing us to probe into the quantum system with minimum effects caused to the system of study [30-35]. Quantum correlations with weak measurement is an emerging research subject that is concerned with the properties of quantum correlations without being savaged by the means of sharp measurement [36-38]. Despite the fact that the correlation dynamics has been broadly studied in open quantum systems, very few works have been addressed the measurements, and the problem is still open for the effect of the environment on quantum discord with weak measurement. In order to understand the role played by measurements, it is therefore desirable to compare the correlation dynamics with the case of strong (projective) and weak measurement. In particular, we want to know whether it is possible to occur a sudden change phenomenon under weak measurement. In this paper, we will examine the dynamics of quantum discord of two qubits under weak measurements. The effect of weak measurements on sudden change of quantum discord will be explored. Further, we will examine whether or not weak measurement change the transition time of the quantum discord.

## II. QUANTUM DISCORD AND WEAK MEASUREMENT

The standard quantum discord is captured by the projective measurements [39]. The quantum discord for a bipartite quantum state $\rho_{AB}$ with a local complete projective measurement $\{\Pi_k^B\}$ on system $B$ is defined as the difference between the total correlation and the classical correlation by the following expression

$$D(\rho_{AB}) = I(\rho_{AB}) - C(\rho_{AB}) \qquad (1)$$

Here, the total correlation between two subsystems $A$ and $B$ of the bipartite quantum system $\rho_{AB}$ is measured by quantum mutual information

$$I(\rho_{AB}) = S(\rho_A) + S(\rho_B) - S(\rho_{AB}) \qquad (2)$$

and $C(\rho_{AB})$ is the classical correlation between the two subsystems. As discussed in Ref. [40], the classical correlation is given by

$$C(\rho_{AB}) = S(\rho_A) - \min_{\{\Pi_k^B\}}[S(\rho_{AB}|\{\Pi_k^B\})] \qquad (3)$$

with the minimization being over all projective-value measurements performed on the subsystem $B$, where $S(\rho_j)$ is the entropy of the quantum system, which is given by the von Neumann entropy

$$S(\rho_j) = -Tr_j(\rho_j \log_2 \rho_j) = -\sum_i \lambda_j^i \log_2 \lambda_j^i \qquad (4)$$

The index $j$ labels either the subsystem $A(B)$ or the total system, and $\{\lambda_j^i\}$ are the nonzero eigenvalues of the quantum state $\rho_j$. Then, $\rho_B$ is the reduced density matrix for the part $B$ and $\rho_{A|k} = Tr_B[(I_A \otimes \Pi_k^B)\rho_{AB}(I_A \otimes \Pi_k^B)]/p_k$, $p_k = Tr_{AB}[(I_A \otimes \Pi_k^B)\rho_{AB}(I_A \otimes \Pi_k^B)]$. The quantum discord is thus defined what we call as the normal quantum discord

$$D(\rho_{AB}) = S(\rho_B) - S(\rho_{AB}) + \min_{\{\Pi_k^B\}} S(\rho_{A|k}) \qquad (5)$$

Indeed, we may extend this definition directly to the quantum discord with weak measurement, called as super quantum discord (SQD) by Singh and Pati [37], by using the notion of weak measurement operators. The theory of weak measurements can be formulated in terms of dichotomic measurement operators as done by Oreshkov and Brun [41]. The weak measurement operators have the following form

$$P(\pm x) = \sqrt{\frac{1 \mp \tanh x}{2}} \Pi_0 + \sqrt{\frac{1 \pm \tanh x}{2}} \Pi_1 \qquad (6)$$

where $x \in R$ is the strength parameter of measurement, and if $x = \varepsilon$, where $|\varepsilon| \ll 1$, the quantum measurement is weak in strength. $\Pi_0$ and $\Pi_1$ are orthogonal projectors that satisfy $\Pi_0 + \Pi_1 = I$. It is easy to find that in the strong measurement limit we have the projective measurement operators, i.e., $\lim_{x \to \infty} P(+x) = \Pi_1$ and $\lim_{x \to \infty} P(-x) = \Pi_0$.

The proposed quantifier for the super quantum discord performed on the subsystem $B$ is given by

$$D_w(\rho_{AB}) = S(\rho_B) - S(\rho_{AB}) + \min_{\{\Pi_k^B\}} S_w(A|P^B(x)) \qquad (7)$$

where the weak quantum conditional entropy can be expressed as

$$S_w(A|P^B(x)) = P(+x)S(\rho_{A|P^B(x)}) + P(-x)S(\rho_{A|P^B(-x)}) \qquad (8)$$

with

$$P(\pm x) = Tr_{AB}[(I_A \otimes P^B(\pm x))\rho_{AB}(I_A \otimes P^B(\pm x))] \quad (9)$$

and the reduced state of the subsystem *A*, after the measurement, can then written as

$$\rho_{A|P^B(\pm x)} = \frac{Tr_B[(I_A \otimes P^B(\pm x))\rho_{AB}(I_A \otimes P^B(\pm x))]}{Tr_{AB}[(I_A \otimes P^B(\pm x))\rho_{AB}(I_A \otimes P^B(\pm x))]} \quad (10)$$

where $I_A$ is the identity operator on the Hilbert space $H_A$, and $P^B(\pm x)$ is the weak measurement operator performed on subsystem *B*.

## III. DYNAMICS OF QUANTUM CORRELATIONS UNDER LOCAL NON-DISSIPATIVE CHANNELS

Two-qubit systems are primary building blocks for encoding correlations via quantum systems. In this section, we discuss the correlation dynamics of two qubits subjects to non-dissipative decoherence processes. In order to examine the peculiar aspect of dynamics of quantum discord between two qubits we use the following approach. We first present our results related to the dynamics of quantum discord for two qubits under local dephasing channels. Then, we analyze possible sudden transitions between different measurement bases, which are used in the calculation of quantum discord, taking place during the time evolution of the system.

Given an initial state for two qubits $\rho(0)$, its evolution under local environments can be written compactly as

$$\rho(t) = \sum_{i,j}(E_i \otimes E_j)\rho(0)(E_i \otimes E_j)^\dagger \quad (11)$$

where $\{E_k\}$ is the set of Kraus operators associated with the decohering process of a single qubit and trace preserving condition $\sum_k E_k^+ E_k = I$ for all $t$.

We consider the case of two qubits under local non-dissipative channels. In this case, decoherence occurs without altering the energy of the system. More specifically, we focus on the Pauli maps (phase flip, bit flip, and bit-phase flip channels) [42].

$$\text{Phase flip:} \quad E_1 = \sqrt{1-p/2}\,I, \quad E_2 = \sqrt{p/2}\,\sigma_3 \quad (12)$$

$$\text{Bit flip:} \quad E_1 = \sqrt{1-p/2}\,I, \quad E_2 = \sqrt{p/2}\,\sigma_1 \quad (13)$$

$$\text{Bit-phase flip:} \quad E_1 = \sqrt{1-p/2}\,I, \quad E_2 = \sqrt{p/2}\,\sigma_2 \quad (14)$$

and the explicit time dependence of the dephasing factor is $p = 1 - \exp(-\gamma t)$ with $\gamma$ being the dephasing rate. We assume here that the local noise channels act on both qubits in the same way with identical dephasing factors. In the following we will consider only phase flip channel as noise model, since the evolutions of the state $\rho(t)$ under bit flip and bit-phase flip channel are symmetric with that of phase flip channel.

Let us assume that the qubits, *A* and *B*, are prepared in Bell diagonal states

$$\rho_{AB} = \frac{1}{4}\left(I_{AB} + \sum_{i=1}^{3} c_i \sigma_i^A \otimes \sigma_i^B\right) \qquad (15)$$

where $\sigma_i^{A(B)}$ is the $i$th component of the standard Pauli operator acting on the $A(B)$ subspace, and $c_i$ is a real number such that $0 \leq |c_i| \leq 1$ for every $i$, and $I_{AB}$ the identity operator of the total system.

We now start our analysis by presenting the dynamics of the quantum discord under local phase flip channel, described by the initial state parameters $c_1(0)$, $c_2(0)$, and $c_3(0)$. According to Eq. (12), the time dependent state parameters are $c_1(t) = \exp(-2\gamma t)c_1(0)$, $c_2(t) = \exp(-2\gamma t)c_2(0)$, and $c_3(t) = c_3(0)$. The evolution of the classical correlations $C$ and the normal quantum discord $D$ are calculated following Ref. [43], and the super quantum discord $D_w$ is calculated following appendix (1). For the other two Pauli channels, we can see the same behavior, except that the relations between the three components of state (15) are exchange [21].

Without loss of generality, in figures 1, 2 and 3, the measures $C$, $D$ and $D_w$ are plotted as a function of scaled time $\gamma t$ for $c_1(0) = 1$, $c_2(0) = -c_3(0)$, $c_3(0) = 0.6$, and different values of $x$. We can find in all figures that the SQD is always larger than the normal quantum discord, which is consistent with the result given by unrestricted measurements [44].

In Figure 1, we plot the time evolution of the normal quantum discord, the super quantum discord and classical correlations for the strength parameter $x = 0.5$. It can be straightforwardly seen that for the normal quantum discord and classical correlations the phenomenon of the sudden change appears in this case. This consequence is in agreement with the results which is obtained by Ref. [22]. However, we should note that for the SQD such a sudden change in behavior dose not occurs in this situation. It is found that the $D_w$ exhibits a monotonically decaying behavior.

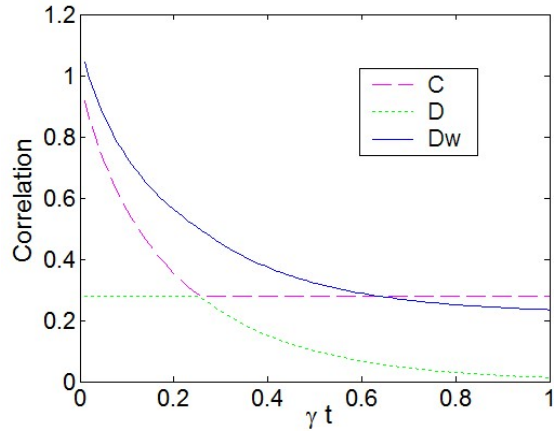

Figure 1 (color online). Dynamics of super quantum discord (blue solid line), classical correlations (pink dashed line), and normal quantum discord (green dotted line) as a function of $\gamma t$ for $C_1(0) = 1$, $C_2(0) = -0.6$, $C_3(0) = 0.6$, and $x = 0.5$.

In order to show the influence of measurements on the behavior of the discord of the two qubits, we can observe the figure 2 and 3 for the strength parameter $x = 1$ and 2, respectively. We can see clearly from figure 2 that the sudden change for $D_w$ appears when $x = 1$. Moreover, we notice in figure 3 that when $x$ increase the numerical value of the $D_w$ remains close to the $D$, but always a little higher. In fact, we have also noted that for $x = 0$, the value of the SQD is equal to the normal quantum discord. Furthermore, from figure 2 and 3, it is not difficult to find that the sudden-change point of $D_w$ is same as $D$ and $C$ for the same parameters. It is shown that the measurement won't change the transition time of the quantum discord. Our analysis shows that quantum discord might exhibit a sudden change only for the projective measurement. Such super quantum discord suggests an important fact that the sudden change does not arise from interaction between the system and environment. And we believe that the environment is not the main role of the sudden change, but the quantum correlation itself is.

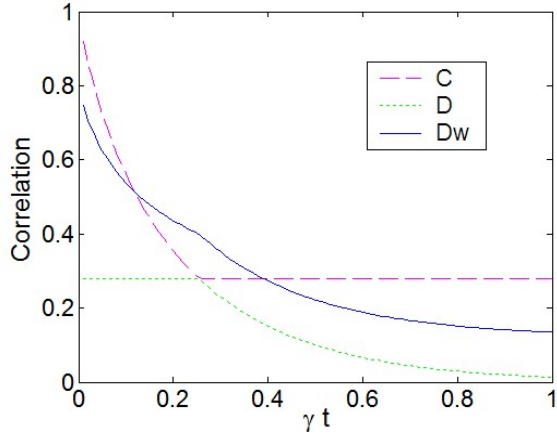

Figure 2 (color online). Same as Figure 1 (color online), but for $x = 1$.

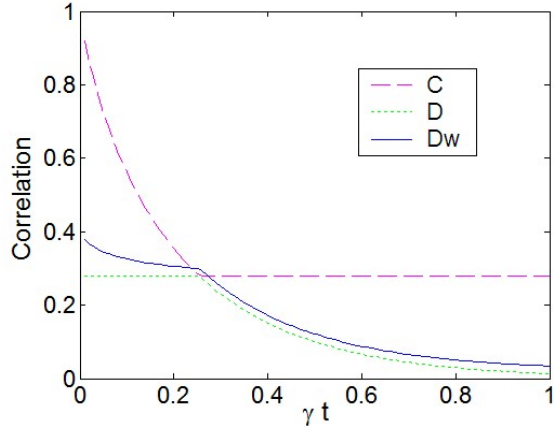

Figure 3 (color online). Same as Figure 1 (color online), but for $x = 2$.

## IV. CONCLUSION

In this paper, we investigate the phenomenon of sudden change of quantum discord for a system of two qubits that are independently interacting with dephasing environments. Instead of choosing projective measurements, we consider weak measurements which allows us to probe into quantum world with minimum effects caused to the system of study. Our results show that the sudden change is intimately related to measurements. We find that the sudden change phenomenon does not appear when the measurement is weak. Quantum discord might exhibit a sudden change only for strong (projective) measurement. This demonstrates that these results contradict the earlier conjecture that different measures of quantum discord give only minor differences. Despite there is still considerable controversy over the role of quantum discord in quantum computation, their dynamics under the action of local non-dissipative environment seems to be, in the present context, intimately relation to measurements. Further studies in this direction might help to elucidate this subtle connection between measurements and quantum correlations, and the role of quantum correlations in quantum computation. We also expect that this study can help to better understand the physical mechanisms of the sudden change, and to simulate novel endeavors in the comprehension and exploitation of quantum discord and measurements.

## ACKNOWLEDGMENTS

We thank Tiangong University 2017 degree and graduate education reform project, Project No. Y20170702.

## APPENDIX

For the Bell diagonal states (15), we have the super quantum discord [45]

$$D_w = -\frac{1-ctanhx}{2}\log_2(1-ctanhx) - \frac{1+ctanhx}{2}\log_2(1+ctanhx)$$
$$+\frac{1}{4}[(1-c_1-c_2-c_3)\log_2(1-c_1-c_2-c_3)$$
$$+(1-c_1+c_2+c_3)\log_2(1-c_1+c_2+c_3)$$
$$+(1+c_1-c_2+c_3)\log_2(1+c_1-c_2+c_3)$$
$$+(1+c_1+c_2-c_3)\log_2(1+c_1+c_2-c_3)] \quad (1)$$

where $c = max\{|c_1|,|c_2|,|c_3|\}$.